\documentclass[%
 reprint,
 amsmath,amssymb,
 aps,
prb,
 longbibliography,
 lengthcheck,%
]{revtex4-1}

\usepackage{graphicx}
\usepackage{dcolumn}
\usepackage{bm}
\usepackage{hyperref}

\begin{document}


\title{Quantum Interference of Impurity Bound States in Bi$_{2}$Sr$_{2}$Ca(Cu$_{1-x}$Zn$_{x}$)$_{2}$O$_{8+\delta}$ Probed by Scanning Tunneling Spectroscopy}

\author{Tadashi Machida$^{1,*}$}
\author{Takuya Kato$^{2}$, Hiroshi Nakamura$^{2}$, Masaki Fujimoto$^{2}$, Takashi Mochiku$^{1}$, Shuuichi Ooi$^{1}$, Ajay D. Thakur$^{1}$, Hideaki Sakata$^{2}$}

\author{Kazuto Hirata$^{1}$}

\affiliation{\\$^{1}$Superconducting Materials Center, National Institute for Materials Science, 1-2-1 Sengen, Tsukuba, Ibaraki 305-0047, Japan \\$^{2}$Department of Physics, Tokyo University of Science, 1-3 Kagurazaka, Shinjuku-ku, Tokyo 162-8601, Japan}
\date{\today}

\begin{abstract}
In conventional superconductors, magnetic impurities form an impurity band due to quantum interference of the impurity bound states, leading to suppression of the superconducting transition temperature. Such quantum interference effects can also be expected in $d$-wave superconductors. Here, we use scanning tunneling microscopy to investigate the effect of multiple non-magnetic impurities on the local electronic structure of the high-temperature superconductor Bi$_{2}$Sr$_{2}$Ca(Cu$_{1-x}$Zn$_{x}$)$_{2}$O$_{8+\delta}$. We find several fingerprints of quantum interference of the impurity bound states including: (i) a two-dimensional modulation of local density-of-states with a period of approximately 5.4 \AA\ along the $a$- and $b$-axes, which is indicative of the $d$-wave superconducting nature of the cuprates; (ii) abrupt spatial variations of the impurity bound state energy; (iii)an appearance of positive energy states; (iv) a split of the impurity bound state. All of these findings provide important insight into how the impurity band in $d$-wave superconductors is formed.
\end{abstract}

\pacs{Valid PACS appear here}
\maketitle
\section{Introduction}
Impurity atoms play an important role in solid state physics, manifested for example by the significant doping dependence of the electrical conductivity in semiconductors.
The effect of impurities is also crucial in superconductors.
In conventional superconductors, magnetic impurities act as pair breakers and thus give rise to a suppression of the superconducting transition temperature ($T_{\mathrm{c}}$), which therefore depends on the impurity concentration \cite{Abrikosov,KMachida}.
When a number of magnetic impurities are present, the wave functions of electrons bound at each impurity potential interfere with each other, leading to the formation of an impurity band with a width that is intimately connected to the reduction of $T_{\mathrm{c}}$ \cite{Shiba,Zittartz}.
\\
\begin{figure}[htb]
\begin{center}
\includegraphics[width=7cm]{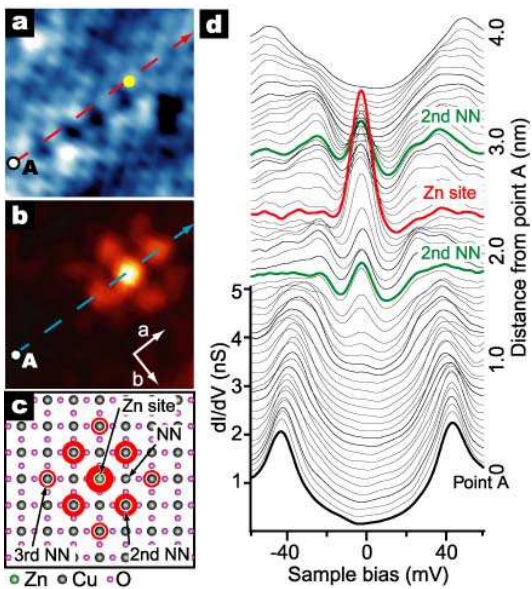}
\end{center}
\caption{
(Color online)
Typical topographic STM image (field of view 3.5 $\times$ 3.5 nm$^2$) of a BiO plane of Zn-doped Bi2212; the Bi atoms are visible as bright spots. The image was taken at a sample bias voltage of $V_{\mathrm{B}}$ = +300 mV and a tunneling current of $I_{\mathrm{set}}$ = 400 pA. (b) Conductance $G(\boldsymbol{r}, V_{\mathrm{B}})$ map at a bias voltage of $V_{\mathrm{B}}$ = -3 mV for the same field of view as in (a). The bright central spot in (b) coincides with the Zn site indicated by the yellow circle in (a).
(c) Schematic representation of the CuO$_{2}$ plane. The intensity maxima (bright spots) in the conductance map of (b) correspond to the Zn atom, second nearest-neighbor (2nd NN) and third nearest-neighbor (3rd NN) sites, as indicated by the red (bold) circles in (c). Local minima in conductance are found at the nearest-neighbor (NN) sites. (d) Evolution of tunneling spectra taken along the line indicated in (a) and (b) (corresponding to the $a$-axis or gap node direction), at different distances from position "A". Only one peak appears, corresponding to the impurity bound state at -3 mV. The peak energy does not depend on the measurement position.
}
\label{Fig_Single}
\end{figure}
\indent
In $d$-wave superconductors such as the cuprates, both magnetic and non-magnetic impurities suppress the bulk $T_{\mathrm{c}}$ \cite{Fukuzumi,Kluge} and modify the local electronic structure \cite{Yazdani,Pan_1,Hudson_1,Hudson_2,Chatterjee,Kambara,TMachida_1}.
The local effect of such impurities has been studied using scanning tunneling spectroscopy (STS), revealing that impurity bound states are created within the superconducting gap, which is (not) suppressed around non-magnetic (magnetic) impurity, in the local density-of-states (LDOS).
For a \textit{single} non-magnetic impurity such as zinc (Zn) \cite{Pan_1,Hudson_2,Kambara,TMachida_1}, the bound state lies close to the $E_{\mathrm{F}}$ as a near-zero-energy peak (NZEP) and the superconducting gap value is suppressed at Zn site.
In real space, the bound state extends into the region surrounding the impurity site and adopts $d$-wave symmetry, referred to as a "cross-shaped pattern".
This pattern is characterized by maximum intensity at the impurity site and secondary maxima at the four second-nearest-neighbor sites \cite{Pan_1,Hudson_2,Kambara,TMachida_1}.
\\
\indent
When a large number of non-magnetic impurities are present in a $d$-wave superconductor, it is theoretically anticipated that the $d$-wave symmetric bound states associated with each impurity site interfere with each other in the same manner as for conventional superconductors \cite{Joynt,Balatsky,Hotta,Alloul,Hussey,Morr,Zhu,Atkinson,Andersen_1,Andersen_2,Andersen_3}. Some theoretical studies have predicted that this interference leads to a splitting and shift in energy of the bound state in the LDOS, as well as to formation of the impurity band, that is closely related to the impurity effect on the bulk properties \cite{Morr,Zhu,Atkinson,Andersen_1,Andersen_2,Andersen_3}.
However, despite a great deal of theoretical effort, few experimental studies have investigated the effect of multiple impurities on the LDOS.
\begin{figure*}[t]
\begin{center}
\includegraphics[width=16cm]{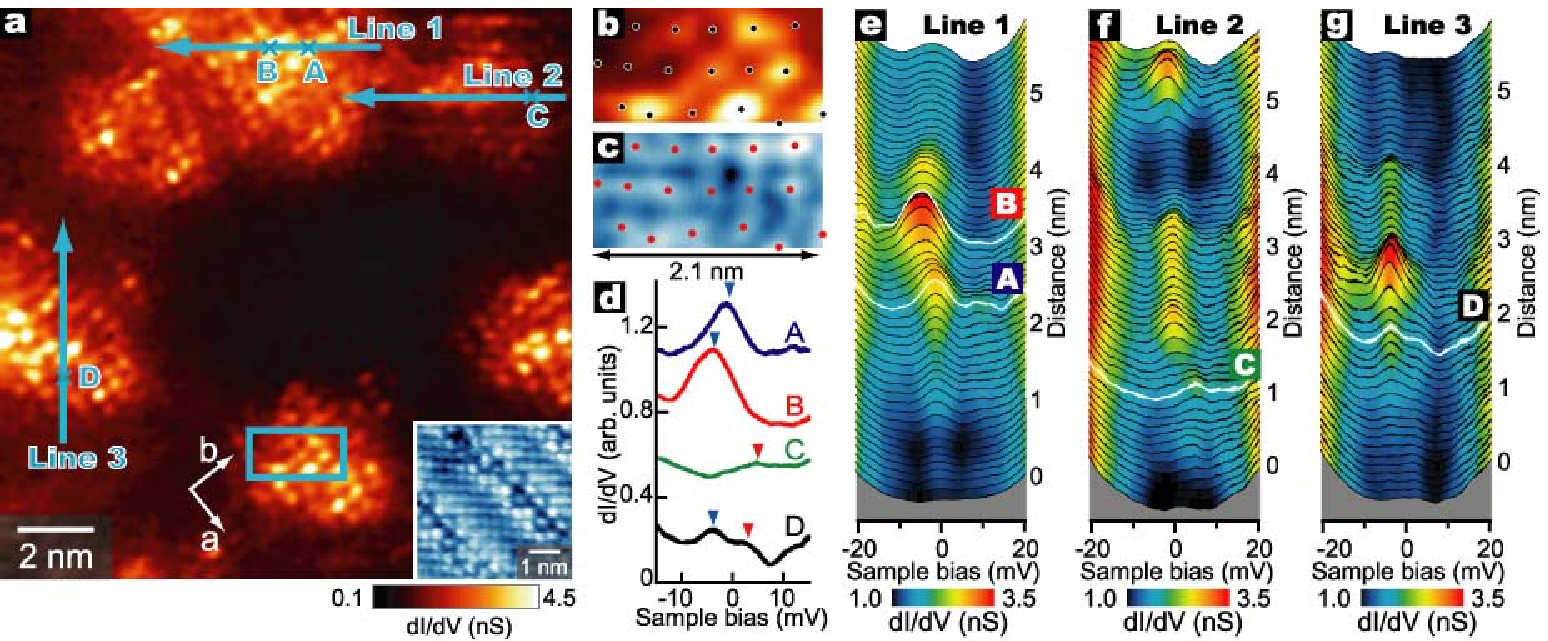}
\end{center}
\caption{
(Color)
(a) Conductance $G(\boldsymbol{r}, V_{\mathrm{B}})$ map (field of view 15 $\times$ 15 nm$^2$) measured at a bias voltage of $V_{\mathrm{B}}$ = -2 mV.
The white arrows indicate the crystal axes determined from the simultaneously obtained topographic image (inset), which was measured with $V_{\mathrm{B}}$ = 150 mV and a tunneling current of $I_{\mathrm{set}}$ = 600 pA.
(b),(c) Magnified conductance map and topographic image for the region outlined by the blue box in (a). The positions of Bi atoms are indicated by black dots in (b) and red dots in (c). (d) Representative tunneling spectra measured at positions "A" to "D" indicated in (a).
For clarity, the spectra at positions "A", "B" and "C" have been shifted upward by 0.75, 0.5, and 0.25, respectively.
Blue and red triangles indicate peak positions at negative and positive energy, respectively. (e)-(g) Tunneling spectra measured at intervals along Lines 1-3 indicated in (a).
The four spectra in (d) are shown in white in (e)-(g).
}
\label{Fig_GMap}
\end{figure*}
\\
\section{Experiment}
Here, we use scanning tunneling microscopy (STM) to study the effect of multiple impurities on the LDOS of single crystal Bi$_{2}$Sr$_{2}$Ca(Cu$_{1-x}$Zn$_{x}$)$_{2}$O$_{8+\delta}$ (Zn-Bi2212, with $T_{\mathrm{c}}$ = 87K for the as-grown sample).
The single crystal used in this study was grown by the floating zone method.
We chose a relatively high Zn concentration of $x \sim$ 0.5\% (the nominal value of $x$ used in the crystal growth was 2\%), which was determined by inductively coupled plasma optical emission spectroscopy performed on the as-grown sample.
The STS measurements were performed in a helium gas environment at 4.2 K on a surface that was atomically clean and flat [see the inset of Fig. \ref{Fig_GMap}(a)]; the surface was prepared by cleaving the sample \textit{in-situ} at 4.2 K in pure helium gas.
We measured the $I$-$V$ characteristics and obtained the tunneling spectra $G(\boldsymbol{r}, V_{\mathrm{B}}) = dI/dV$ by numerical differentiation.
\\
\section{Results and Discussion}
In most regions of the sample probed in this study, we observed \textit{isolated} Zn resonance sites with the characteristic "\textit{cross shaped pattern}", as shown in Fig. \ref{Fig_Single}; the concentration of these sites was approximately 0.45\% per Cu atom (7 $\sim$ 8 counts in a 15 $\times$ 15 nm$^{2}$ field of view), which is comparable to the concentration of Zn in the bulk.
\\
\indent
According to theoretical work \cite{Morr,Zhu,Atkinson}, when the distance between two Zn atoms becomes shorter than 10$a_{0}$ ($a_{0}$ is Cu-O-Cu distance), a spatial modulation of the LDOS near $E_{\mathrm{F}}$ should appear in the vicinity of the Zn atoms as a result of interference between the bound states.
Indeed, we found some regions of the sample where such a LDOS modulation partially appears, as shown by the $G(\boldsymbol{r}, -2\ \mathrm{mV})$ map for a 15 $\times$ 15 nm$^{2}$ field of view in Fig. \ref{Fig_GMap}(a).
Several bright islands are apparent, in which a two-dimensional (2D) LDOS modulation along the $a$- and $b$-axes is visible with a period of approximately 5.4 \AA\ ($\sqrt{2}\times a_{0}$); this is shown more clearly in Figs. \ref{Fig_GMap}(b) and (c).
Figures \ref{Fig_GMap}(e)-(g) show series of tunneling spectra measured at intervals along the three lines indicated in Fig. \ref{Fig_GMap}(a), which all pass through the bright islands. All spectra taken in the bright islands exhibit a NZEP corresponding to the impurity bound state.
This implies that the 2D LDOS modulation is strongly related to the impurity bound state. However, the observed spatial pattern is a \textit{lattice-shaped pattern}, which is significantly different to the \textit{cross-shaped pattern} observed in the vicinity of an "\textit{isolated single}" impurity as shown in Fig. \ref{Fig_Single}(b).
Therefore, it is conceivable that two or more impurities are present in the bright islands in the $G(\boldsymbol{r}, -2 \mathrm{mV})$ map, although the exact positions of these impurities could not be identified.
If we assume that there are at least two impurity atoms in each bright island in Fig. \ref{Fig_GMap}(a), there are more than 12 impurity atoms in this field of view, corresponding to a concentration of at least 0.8\% per Cu atom.
\\
\indent
In addition to these \textit{lattice-shaped pattern}, we observed several more features that are significantly different to those present near an "\textit{isolated single}" impurity, as illustrated by the spectral surveys on the 2D LDOS modulation in Figs. \ref{Fig_GMap}(e)-(g). First, the energy of the NZEP exhibits a pronounced spatial variation. For example, in Fig. \ref{Fig_GMap}(e) the peak energy drastically shifts from -1 meV at site A to -4 meV at site B, despite the relatively short distance ($\sim$ 7 \AA) between these sites.
Second, positive energy states appear, as shown in Fig. \ref{Fig_GMap}(f) where the peak energy abruptly shifts from -2 meV to +5 meV at site C.
Third, a splitting of the impurity bound state is observed. For example, the spectrum at the bottom of Fig. \ref{Fig_GMap}(g) has only one broad peak at positive energy.
As one moves along Line 3 to site D, a negative energy peak gradually appears at -4 meV.
At site D, two peaks at -4 meV and +3 meV are observed, and are shown more clearly in Fig. \ref{Fig_GMap}(d).
\\
\indent
These three peculiar phenomena in the spectral surveys can be explained as a result of quantum interference among the impurity bound states by comparison with theoretical calculations \cite{Morr,Zhu,Andersen_1,Andersen_2,Andersen_3}.
The simplest situation for which this interference effect can theoretically be discussed is the two-impurity case. It is predicted that at most four bound states will be generated (two states at negative energy and two states at positive energy) \cite{Morr,Zhu,Andersen_1,Andersen_2,Andersen_3}.
According to these calculations, the spectral weights and the energies of these states depend on the measurement position and the distance between the impurities, respectively. Furthermore, it has been predicted that (i) an abrupt change in the bound state energy, (ii) an appearance of positive energy states, and (iii) a split of the bound state can occur as a result of quantum interference when the impurity atoms are found in a particular configuration; these are all phenomena observed in our current study.
Although a quantitative comparison of our experimental data with the theoretical calculations is not possible due to the uncertainty in the impurity positions, the observed features qualitatively agree with the calculations.
Because the three features ascribed to the quantum interference effect occur on the 2D LDOS modulation, we expect that the modulation also originates from the quantum interference.

\begin{figure*}[tb]
\begin{center}
\includegraphics[width=16cm]{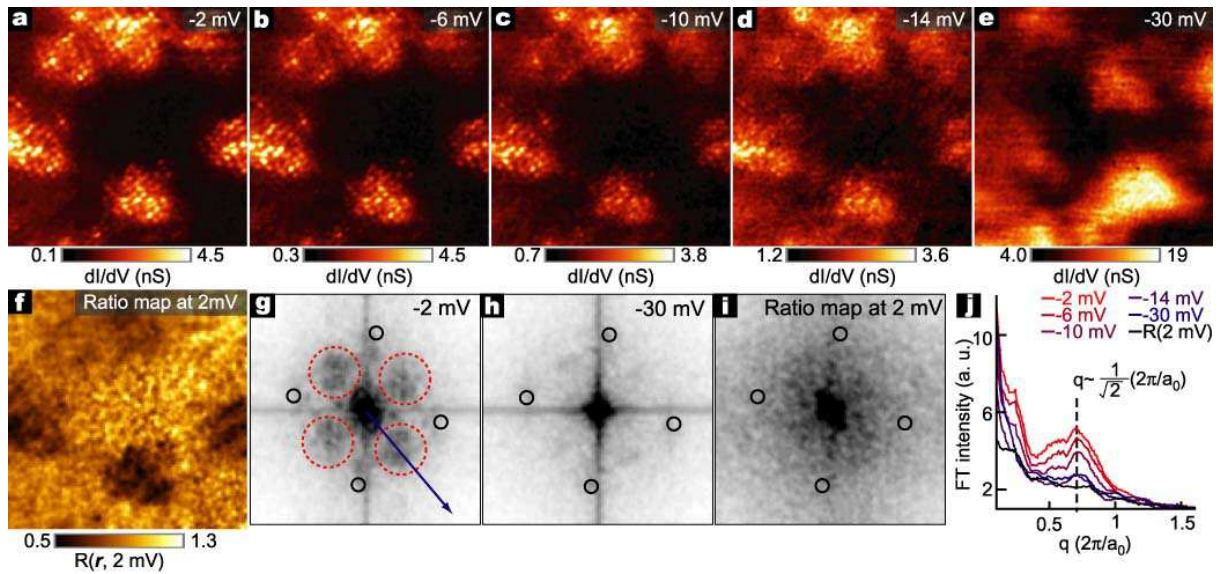}
\end{center}
\caption{
(Color online)
(a)-(e) Conductance $G(\boldsymbol{r}, V_{\mathrm{B}})$ maps at $V_{\mathrm{B}}$ = -2, -6, -10, -14, and -30 mV for the same field of view as in Fig. \ref{Fig_GMap}(a).
(f) Conductance ratio map $R(\boldsymbol{r}, V_{\mathrm{B}})$ = $G(\boldsymbol{r}, +V_{\mathrm{B}})$/ $G(\boldsymbol{r}, -V_{\mathrm{B}})$ at $V_{\mathrm{B}}$ = 2 mV.
(g)-(i) Fourier transform images corresponding to maps (a), (e), and (f), respectively. The four black circles indicate the Fourier spots of the atomic lattice.
In (g), four broad spots corresponding to the 2D modulation are indicated by red dashed circles; no such spots are visible in (h) and (i).
(j) FT intensity taken along the cut indicated by the blue arrow in (g) for various values of $V_{\mathrm{B}}$.
The black line represents the corresponding line profile in the ratio map at $V_{\mathrm{B}}$ = 2 mV.
}
\label{Fig_EDep}
\end{figure*}

\indent
In order to confirm the origin of the 2D LDOS modulation, we investigated the energy evolution of the $G(\boldsymbol{r}, V_{\mathrm{B}})$ map and performed Fourier analyses as shown in Fig. \ref{Fig_EDep}.
In the Fourier transform (FT) image of the $G(\boldsymbol{r}, -2 \mathrm{mV})$ map shown in Fig. \ref{Fig_EDep}(g), the four broad spots indicated by red dashed circles correspond to the 2D LDOS modulation. In the cut taken along the blue line in this FT image, a broad peak appears at $q \sim$ (1/$\sqrt{2}$)$(2\pi/a_{0})$, as indicated in Fig. \ref{Fig_EDep}(j). This is consistent with the real-space analysis regarding the modulation.
As the energy is increased up to $|E| = e|V_{\mathrm{B}}| = 10\ \mathrm{meV}$, the 2D LDOS modulation becomes less well defined and is no longer visible at energies above 14 meV. Instead, the conductance map then shows an inhomogeneous pattern reflecting the inhomogeneous superconducting gap distribution, which commonly exists in cuprate superconductors \cite{Pan_2,Lang,McElroy_2,Gomes,Pasupathy,Kohsaka_2,TMachida_2,Kato}.
The period of the 2D LDOS modulation does not change with increasing energy within our $q$-space resolution ($\sim$ 0.025 $\times $ $2\mathrm{\pi}/a_{0}$), as shown in Fig. 3(j).
In addition to this non-dispersive nature, the amplitude of the LDOS modulation is strongly suppressed by taking the ratio between the conductance at a positive and a negative bias voltage [$R(\boldsymbol{r}, V_{\mathrm{B}})=G(\boldsymbol{r}, +V_{\mathrm{B}})/G(\boldsymbol{r}, -V_{\mathrm{B}})$], as shown in Fig. \ref{Fig_EDep}(f) and the corresponding FT image in Fig. \ref{Fig_EDep}(i).
\\
\indent
From these results, several possible explanations for the origin of the observed LDOS modulation can be excluded.
First, the non-energy-dispersive nature of the LDOS modulation and the suppression of amplitude in the $R(\boldsymbol{r}, V_{\mathrm{B}})$ map are inconsistent with the properties of \textit{quasiparticle interference} (QPI), which has been observed for several cuprate superconductors \cite{Hoffman_2,McElroy_1,Hanaguri,Kohsaka_1}, since the amplitude of the QPI pattern should be enhanced in the $R(\boldsymbol{r}, V_{\mathrm{B}})$ map \cite{Hanaguri,Fujita}.
Furthermore, the observed 2D LDOS modulation is completely different to the well-known non-energy-dispersive LDOS modulation with a period of $\sim$ 4$a_{0}$$\times$4$a_{0}$ (so-called "\textit{checkerboard modulation}") and $\sim$ 4/3$a_{0}$$\times$4/3$a_{0}$ (so-called "\textit{nano-stripe}") modulation that are observed in both underdoped cuprates \cite{TMachida_2,Hanaguri_2,Kohsaka_3,Wise} and around vortex cores \cite{Matsuba,Hoffman}.
In our case the observed modulation is aligned along the $a$- and $b$-axes, which are tilted by 45$^{\circ}$ from the direction of the checkerboard and nano-stripe modulations.

The features that we observe agree rather well with the predictions of calculations regarding the quantum interference effect of randomly distributed impurities \cite{Atkinson}.
These calculations indicate that 2D LDOS modulation can partially appear along the $a$- and $b$-axes with a period of 5.4 \AA\ as a result of the interference of impurity bound states when the impurity concentration is 0.5\% per Cu atom, which is almost the same as that in our sample.
Thus, it is plausible that the 2D LDOS modulation and the three peculiar phenomena observed in the spectral surveys reflect the quantum interference effect.

\section{Summary}
In summary, we have investigated the effect of multiple impurities on the LDOS in Zn-Bi2212 using scanning tunneling spectroscopy.
Our results reveal several peculiar features: (i) the 2D LDOS modulation with a period of approximately 5.4 \AA\ along $a$- and $b$-axis; (ii) the abrupt change in the energy of the impurity bound states with position; (iii) the existence of positive energy states; (iv) the splitting of the impurity bound state.
All of these features have been predicted in several theoretical studies focused on the effect of quantum interference on the impurity bound states.
Due to the good qualitative agreement between our results and these theoretical calculations, we conclude that the observed features are fingerprints of the quantum interference of impurity bound states.
Furthermore, one may expect that the accumulation of these states at several energies induced by the interference creates the impurity band that is intimately related to the bulk transport properties.
Thus, the quantum interference of impurity bound states observed in this study plays a crucial role in understanding how the microscopic impurity effect is linked to macroscopic properties such as the suppression of $T_{\mathrm{c}}$.

\nocite{*}
\thebibliography{99}
\bibitem{Abrikosov} A. A. Abrikosov and L. P. Gor'kov, Sov. Phys. JETP \textbf{12}, 1243 (1961).
\bibitem{KMachida} K. Machida and F. Shibata, Prog. Theor. Phys. \textbf{47}, 1817 (1972).

\bibitem{Shiba} H. Shiba, Prog. Theor. Phys. \textbf{40}, 435 (1968).
\bibitem{Zittartz} J. Zittartz, A. Bringer, and E. M-Hartmann, Solid State Commun. \textbf{10}, 513 (1972).

\bibitem{Fukuzumi} Y. Fukuzumi, K. Mizuhashi, K. Takenaka, and S. Uchida, Phys. Rev. Lett. \textbf{76}, 684 (1996).
\bibitem{Kluge} T. Kluge, Y. Koike, A. Fujiwara, M. Kato, T. Noji, and Y. Saito, Phys. Rev. B \textbf{52}, R727 (1995).

\bibitem{Yazdani} A.Yazdani, C.M. Howald, C. P. Lutz, A. Kapitulnik, and D. M. Eigler, Phys. Rev. Lett. \textbf{83}, 176 (1999).
\bibitem{Pan_1} S. H. Pan, E. W. Hudson, K. M. Lang, H. Eisaki, S. Uchida, and J. C. Davis, Nature (London) \textbf{403}, 746 (2000).
\bibitem{Hudson_1} E. W. Hudson, S. H. Pan, A. K. Gupta, K.-W. Ng, and J. C. Davis, Science \textbf{285}, 88 (1999).
\bibitem{Hudson_2} E. W. Hudson, K. M. Lang, V. Madhavan, S. H. Pan, H. Eisaki, S. Uchida, and J. C. Davis, Nature (London) \textbf{411}, 920 (2001).
\bibitem{Chatterjee} K. Chatterjee, M. C. Boyer, W. D. Wise, T. Kondo, T. Takeuchi, H. Ikuta, and E. W. Hudson, Nature Phys. \textbf{4}, 108 (2008).
\bibitem{Kambara} H. Kambara, Y. Niimi, M. Ishikado, S. Uchida, and H. Fukuyama, Phys. Rev. B \textbf{76}, 052506 (2007).
\bibitem{TMachida_1} T. Machida, T. Kato, H. Nakamura, M. Fujimoto, T. Mochiku, S. Ooi, A. D. Thakur, H. Sakata, and K. Hirata, Phys. Rev. B \textbf{82} 180507(R) (2010).

\bibitem{Joynt} R. Joynt, J. Low Temp. Phys. \textbf{109}, 811 (1997).
\bibitem{Balatsky} A. V. Balatsky, I. Vekhter, and J.-X. Zhu, Rev. Mod. Phys. \textbf{78}, 373 (2006).
\bibitem{Hotta} T. Hotta, J. Phys. Soc. Jpn. \textbf{62}, 274 (1993).
\bibitem{Alloul} H. Alloul \textit{et al.}, Rev. Mod. Phys. \textbf{81}, 45 (2009).
\bibitem{Hussey} N. E. Hussey, Adv. Phys. \textbf{51}, 1685 (2002).
\bibitem{Morr} D. K. Morr and N. A. Stavropoulos, Phys. Rev. B \textbf{66}, 140508(R) (2002).
\bibitem{Zhu} L. Zhu, W. A. Atkinson, and P. J. Hirschfeld, Phys. Rev. B \textbf{67}, 094508 (2003).
\bibitem{Atkinson} W. A. Atkinson, P. J. Hirschfeld, and L. Zhu, Phys. Rev. B \textbf{68}, 054501 (2003).
\bibitem{Andersen_1} B. M. Andersen and P. Hedeg\aa rd, Phys. Rev. B \textbf{67}, 172505 (2003).
\bibitem{Andersen_2} B. M. Andersen, Phys. Rev. B \textbf{68}, 094518 (2003).
\bibitem{Andersen_3} B. M. Andersen, A. Melikyan, T. S. Nunner, and P. J. Hirschfeld, Phys. Rev. Lett. \textbf{96}, 097004 (2006).
\bibitem{Pan_2} S. H. Pan, J. P. O'Neal, R. L. Badzey, C. Chamon, H. Ding, J. R. Engelbrecht, Z. Wang, H. Eisaki, S. Uchida, A. K. Gupta, K.-W. Ng, E. W. Hudson, K. M. Lang, and J. C. Davis, Nature (London) \textbf{413}, 282 (2001).
\bibitem{Lang} K. M. Lang, V. Madhavan, J. E. Hoffman, E. W. Hudson, H. Eisaki, S. Uchida, and J. C. Davis, Nature (London) \textbf{415}, 412 (2002).
\bibitem{McElroy_2} K. McElroy, Jinho Lee, J. A. Slezak, D.-H. Lee, H. Eisaki, S. Uchida, and J. C. Davis, Science \textbf{309}, 1048 (2005).
\bibitem{Gomes} K. K. Gomes, A. N. Pasupathy, A. Pushp, S. Ono, Y. Ando, and A. Yazdani, Nature (London) \textbf{447}, 569 (2007).
\bibitem{Pasupathy} A. N. Pasupathy, A. Pushp, K. K. Gomes, C. V. Parker, J. Wen, Z. Xu, G. Gu, S. Ono, Y. Ando, and A. Yazdani, Science \textbf{320}, 196 (2008).
\bibitem{Kohsaka_2} Y. Kohsaka, K. Iwaya, S. Satow, T. Hanaguri, M. Azuma, M. Takano, and H. Takagi, Phys. Rev. Lett. \textbf{93}, 097004 (2004).
\bibitem{Kato} T. Kato, S. Okitsu, and H. Sakata, Phys. Rev. B \textbf{72}, 144518 (2005).
\bibitem{TMachida_2} T. Machida, Y. Kamijo, K. Harada, T. Noguchi, R. Saito, T. Kato, and H. Sakata, J. Phys. Soc. Jpn. \textbf{75}, 083708
    (2006).

\bibitem{Hoffman_2} J. E. Hoffman, K. McElroy, D.-H. Lee, K. M Lang, H. Eisaki, S. Uchida, and J. C. Davis, Science \textbf{297}, 1148 (2002).
\bibitem{McElroy_1} K. McElroy, R. W. Simmonds, J. E. Hoffman, D.-H. Lee, J. Orenstein, H. Eisaki, S. Uchida, and J. C. Davis, Nature (London) \textbf{422}, 592 (2003).
\bibitem{Kohsaka_1} Y. Kohsaka, C. Taylor, P. Wahl, A. Schmidt, Jhinhwan Lee, K. Fujita, J. W. Alldredge, K. McElroy,
Jinho Lee, H. Eisaki, S. Uchida, D.-H. Lee, and J. C. Davista, Nature (London) \textbf{454}, 1072 (2008).
\bibitem{Hanaguri} T. Hanaguri, Y. Kohsaka, J. C. Davis, C. Lupien, I. Yamada, M. Azuma, M. Takano, K. Ohishi, M. Ono, and H. Takagi, Nature Phys. \textbf{3}, 865 (2007).
\bibitem{Fujita} K. Fujita, Ilya Grigorenko, J. Lee, W. Wang, Jian Xin Zhu, J. C. Davis, H. Eisaki, S. Uchida, and Alexander V. Balatsky, Phys. Rev. B \textbf{78}, 054510 (2008).

\bibitem{Hanaguri_2} T. Hanaguri, C. Lupien, Y. Kohsaka, D.-H. Lee, M. Azuma, M. Takano, H. Takagi, and J. C. Davis, Nature \textbf{430}, 1001 (2004).
\bibitem{Kohsaka_3} Y. Kohsaka, C. Taylor, K. Fujita, A. Schmidt, C. Lupien, T. Hanaguri, M. Azuma, M. Takano, H. Eisaki, H. Takagi, S. Uchida, J. C. Davis, Science \textbf{315}, 1380 (2007).
\bibitem{Wise} W. D. Wise, M. C. Boyer, K. Chatterjee, T. Kondo, T. Takeuchi, H. Ikuta, Y. Wang, and E. W. Hudson, Nature Phys. \textbf{4}, 696 (2008).
\bibitem{Matsuba} K. Matsuba, S. Yoshizawa, Y. Mochizuki, T. Mochiku, K. Hirata, and N. Nishida, J. Phys. Soc. Jpn. \textbf{76}, 063704 (2007).
\bibitem{Hoffman} J. E. Hoffman, E. W. Hudson, K. M. Lang, V. Madhavan, H. Eisaki, S. Uchida, J. C. Davis, Science \textbf{295}, 466 (2002).
\end{document}